\documentclass[sigconf]{acmart}
\usepackage{subfigure}
\usepackage{multirow}

\AtBeginDocument{%
  \providecommand\BibTeX{{%
    \normalfont B\kern-0.5em{\scshape i\kern-0.25em b}\kern-0.8em\TeX}}}

\setcopyright{acmcopyright}
\copyrightyear{2019}
\acmYear{2019}

\acmConference[Cascon '19]{Cascon '19: 29th Annual International Conference on Computer Science and Software Engineering}{Nov 04--06, 2019}{Markam, Toronto}
  
\begin{document}

\title{A Fog Computing Framework for Autonomous Driving Assist: Architecture, Experiments, and Challenges}

\author{Muthucumaru Maheswaran}
\affiliation{%
  \institution{McGill University}
  \city{Montreal}
  \state{Quebec}
}
\email{maheswar@cs.mcgill.com}

\author{Tianzi Yang}
\affiliation{%
  \institution{McGill University}
  \city{Montreal}
  \state{Quebec}
}
\email{tianzi.yang@mail.mcgill.ca}

\author{Salman Memon}
\affiliation{%
  \institution{McGill University}
  \city{Montreal}
  \state{Quebec}
}
\email{salman.memon@mail.mcgill.ca}

\begin{abstract}
Autonomous driving is expected to provide a range of far-reaching economic, environmental and safety benefits. In this study, we propose a fog computing based framework to assist autonomous driving.
Our framework relies on overhead views from cameras and data streams from vehicle sensors to create a network of distributed digital twins, called an edge twin, on fog machines. The edge twin will be continuously
updated with the locations of both autonomous and human-piloted vehicles on the road segments.
The vehicle locations will be
harvested from overhead cameras as well as location feeds from the vehicles themselves.
Although the edge twin can make fair road space allocations from a global viewpoint, there is a communication cost (delay) in reaching it from the cameras and vehicular sensors. To address this, we introduce a machine learning forecaster as a part of the edge twin which is responsible for predicting the future location of vehicles.
Lastly, we introduce a box algorithm that will use the forecasted values to create a hazard map for the road segment which would be used by the framework to suggest safe manoeuvres for the autonomous vehicles such as lane changes and accelerations. We present the complete fog computing framework for autonomous driving assist and evaluate
key portions of the proposed framework using simulations based on a real-world dataset of vehicle position traces on a highway.
\end{abstract}


\maketitle

\section{Introduction} 
\label{sec:intro}

Autonomous driving is billed as a major technology disruptor since the invention of the web. Different stakeholders from a variety of technology sectors are investing significant time and money on solving this complex problem and have made significant progress towards a working solution. However, given the scope and complexity of the problem, it remains far from solved. Inspired by another emerging technology - fog computing, this paper brings a novel perspective to the problem. The key idea is to consider an integrated approach to the autonomous driving problem where
instead of each vehicle solving the problem in isolation, a fog powered framework takes a bird's eye view of the situation and offers directives to the cars from that perspective. 

Fog computing pushes the compute resources of
the cloud closer to the edge of the network \cite{bonomi2012fog}. It provides an ideal platform for handling data-intensive applications at the edge closer to the source of data.  We leverage this aspect of fog computing to process feeds from overhead cameras to map out the locations of the vehicles and other road objects at real time and use that information to provide drive assist. 

The fog computing based framework we present here is inspired by digital twin, which is a sophisticated virtual (cyber)
representation of a physical entity~\cite{schleich2017shaping}. 
Our framework would maintain digital twins for road segments in the fog servers and keep them updated with data from cameras observing the road segments. This way, the fog servers would be aware of the happenings in the road segments. 

We call the network of digital twins distributed across the fogs and the underlying middleware substrate governing their operations the edge twin. The edge twin would be responsible for:
\begin{itemize}
\item Engaging a fog server and make it useable to serve requests from autonomous vehicles.
\item Maintaining connectivity between cars and the optimal fog servers such that requests from cars are served with smallest application-level latency.
\item Keeping the system fault-tolerant despite fog failures and camera failures. 
\item 
Facilitating the deployment of applications that solve autonomous driving tasks using data that is harvested in real-time.
\end{itemize}

One of the challenges of keeping the edge twin up-to-date with the latest physical world view is the application-level communication delay in reaching the fog servers from the data sources (i.e., cameras) and vehicles. That is, the edge twin would be running an application task on a time-delayed view of the world. To compensate for the time-delayed view, the edge twin would have a time-shifting module to move the world view to the correct time point. 

The time shifting is achieved by having a machine learning trajectory predictor for the moving vehicles. Using the trajectory predictors, we can forecast the locations at a future time and run the application tasks on that world view. If the forecasts have a minimal error, we can hide the staleness introduced by the time delay in updating the edge twin and processing tasks. 

The edge twin is an ideal host for any application that needs a real-time updated world view such as the auto-drive assist. One of the important tasks of auto-drive assist is road space allocation. We discretize the road space into boxes of a predefined size and determine whether it is safe to allocate them to autonomous vehicles that need a pathway for their journeys. 

There are two problems that the road space allocator needs to solve. First is to determine the free boxes -- boxes that are not occupied by the human-piloted vehicles. Because human-piloted vehicles are not under the edge twin's control, they are free to move in any feasible manner. We use machine learning forecasters to predict their movement and determine boxes that are least likely going to be occupied by them. The second problem is to fairly and efficiently allocate the free boxes among the autonomous cars so that they are able to make timely progress with their journeys without colliding with each other. By consolidating road space allocation at the edge twin, this approach provides a finer way of controlling road space usage (e.g., dealing with traffic congestion).

In the following section, we discuss background concepts such as fogs, digital twins, and mirror world. We present a motivation for fog-based approach for 
autonomous driving assist in Section~\ref{motivation}. The system architecture 
is described in detail in Section~\ref{sec:sysarc}.
The proof-of-concept implementation of portions of the system architecture as
covered by this paper is provided in Section~\ref{impl}. Section~\ref{expr} describes the
results from the trace-driven simulations on the neural networking models. 
Lastly, in Section \ref{sec:related}, we go over related literature pertaining to the use of fog computing, digital twins and machine learning for autonomous driving assistance.

\section{Background Work}\label{sec:background}

\subsection{Fog Computing}

Fog computing is a distributed computing architecture that introduces an intermediate layer of devices between the cloud and end-devices \cite{dastjerdi2016fog}. This essentially extends the compute and storage resource typically provided by the cloud closer to the edge of the network. 
Fog computing has a significant overlap with related technologies such as Mobile Edge Computing (MEC) and Mobile Cloud Computing (MCC), and the terms are sometimes used interchangeably \cite{yi2015survey} \cite{shi2016edge}. 

Fog computing offers numerous benefits which can be boiled down to two categories: an enhanced user experience for end users and enhanced network efficiency \cite{luan2015fog}. End users can experience significantly higher access speeds and lower latency communicating with fog servers rather than the cloud \cite{shi2016edge}. It also minimizes the back-and-forth traffic between the core network and end users, thereby reducing bandwidth usage on these busy links and improving network efficiency\cite{luan2015fog}.  

These characteristics make fog computing an excellent candidate for technologies such as connected and autonomous vehicles which produce large volumes of data at very high rates. Fog computing can allow autonomous vehicles to process large volumes of data at or close to real-time speeds \cite{shi2016edge}. The fog servers can be directly connected to Road Side Units (RSUs) which are fixed infrastructure elements in place to aid autonomous vehicles \cite{datta2015fog}.
Fog computing is also part of the {\em Multi-Access Edge Computing}
paradigm that is integrated into the 5G cellular architecture \cite{mobileedgecomp}. With 
5G, fog servers would be located at the 5G base stations so that performance 
critical portions of a cloud-based application can be hosted in the fog. \cite{giust2018multi} provides an overview of some of the applications for autonomous driving that MEC in 5G would make available. 


\subsection{Digital Twin}
A digital twin is a data abstraction of a natural living or non-living object, where data can be exchanged seamlessly between the physical and the virtual counterparts~\cite{D2,D1}. The abstraction can be complicated or straightforward according to the use case, where the more sophisticated description will determine the precision of the physical object representation~\cite{D3}. Digital twins are the offspring of many disciplines, including IoT, machine learning, predictive data analytics and spatial network graphs. A more advanced version of a digital twin is a predictive twin, where it can model the future status of an object~\cite{D4}. By exploiting the historical data of a digital twin and through continuous learning, a predictive twin can predict the future behavior of the physical object. 

Digital twins offer a range of advantages that make them invaluable to a variety of different technology sectors. The new focus on curating and utilizing data sources in the product design and manufacturing industry has played a major role in improving the quality of digital twins to make them almost synonymous with the physical products \cite{tao2018digital}. Having highly accurate digital representations is also observed to drastically cut costs in the prototype design and testing phases \cite{grieves2017digital}. Moreover, they also contribute to improving the quality of the products by allowing for easier design evaluations, life-cycle estimation and certifications \cite{glaessgen2012digital} 

\subsection{Mirror Worlds}

Mirror Worlds is a much richer concept than digital twin coined by Gelernter~\cite{gelernter1993mirror} in 1990s. It has 
inspired many modern applications such as Google Earth~\cite{mirror_worlds}. Like digital twin a mirror world 
would represent a physical system using software inside a computer. The computer representation will be live and would have history. The physical system can be controlled using the mirror world. Despite the many thought provoking ideas presented in~\cite{gelernter1993mirror}, the overall concept is left hypothetical.

\section{Motivating Scenarios}
\label{motivation}

The primary purpose of this paper is to make a case that a concept like the
Mirror world~\cite{gelernter1993mirror} can immensely benefit a cyber-physical
command and control problem such as autonomous drive assist. In this section, we
describe several example scenarios that highlight the benefits of a mirror world
system. To keep the discussions simple, we use an external server as a stand-in
for the mirror world.

\subsection{Benefits of an Outside-the-Car Observer}

Consider a scenario where a pedestrian is walking towards a street crossing. An
autonomous car approaching the crossing would observe the pedestrian for a very
short period of time (i.e., when the pedestrian comes in the view of its
sensors). It needs to use the information gathered from that brief observation
to decide whether there is a risk of the pedestrian walking into the street.
This would not be sufficient in many cases, so the car needs to make a
conservative decision to slow down. By slowing down, the car retains the ability
to come to a sudden halt if the pedestrian walks into the street crossing and
thus avoid an accident. If the vehicle was able to forecast the next move of the
pedestrian with very high confidence it could avoid unnecessary slow downs.
Suppose we place a fog-based outside-the-car observer (i.e., an AI-enabled
camera) in that segment of the road such that it can watch the happenings all
the time. The autonomous car can consult that observer to obtain a highly
confident prediction of the pedestrian's next move. Because the fog-based observer is continuously watching the space, it can observe the pedestrian over a longer period of time, while also having the luxury to use historical data on how other pedestrians behaved in that space.

To make the observer idea useful, we need to have them deployed in a pervasive
manner, and they must be accessible with ultra-low latency from the vehicles. A
vehicle would join the observer responsible for the road space it is entering at
any given moment. If the observer is not available, the vehicle needs to fall
back to full local decision making, which could slow down the overall traffic
flow.

Because the outside-the-car observer has a bird's-eye view of the traffic
conditions, it can be consulted by a vehicle to know the next move of
the vehicles standing in its lane ahead of it. If the observer can predict that
the vehicles would start moving within a short time with high confidence, the
oncoming vehicle does not need to change lanes.

The vantage point of the outside-the-car observer is independent of the
in-the-car sensing employed by the vehicles. Therefore, obstacles not noticed
using in-the-car sensing could register in the
outside-the-car observer. This diversity in the data perspective can
significantly reduce the risk of not detecting potential hazards for driving
maneuvers. Also, when snow conditions wipe out the lane markings, the
outside-the-car observer should still work because it would be able to locate
the vehicles in its fixed view without lane markings.

\subsection{Information Sharing Through an Outside-the-Car Observer}

Dedicated short-range communication (DSRC)~\cite{dsrc} is part of connected car
technology that can allow cars to quickly share information regarding sudden brakings
or lane changes. There are many valuable road information that is not best
disseminated using a direct approach like DSRC. For instance, pothole
information or changing driveability conditions during wildfires are best
managed through outside-the-car observers. Using the outside-the-car observer
approach, the observer can pre-process the data feeds obtained from the vehicles
or other road objects before passing them along to others. This allows the
observers to contextualize the data feeds and also reduce redundant information
processing at the vehicles. Furthermore, vehicles need not be at the same place at
the same time to exchange information. The observer can store the data feeds and
aggregate them into new feeds depending on the semantics of the data.

The global vantage of fog-based observers could allow them to cognitive processing tasks that extract a higher
level of actionable intelligence than what could be extracted from a single data
feed alone.  For example, we could use the cellular network activity of a user
along with the physical behavior observed by analyzing  the direct data feeds
captured by the observer (e.g., the overhead traffic camera feed) to identify
distracted pedestrian or driver and preemptively react to a hazardous situation.

\subsection{Efficient Use of Road Space}

Traffic congestion is a problem plaguing many cities. An external observer would
provide an ideal platform to implement allocation algorithms for efficiently
dividing road space among competing vehicles. The current road space allocation
is carried out, assuming two traffic models: homogeneous and heterogeneous.
Homogeneous traffic models lead to a lane-based division of road space and
heterogeneous traffic models lead to lane-less
division~\cite{asaithambi2016driving}. Many advanced cities that are
anticipating the introduction of autonomous cars are also encouraging bikes,
e-bikes, and other slow-moving eco-friendly modes of transportation. Therefore,
static partitioning of road space into lanes needs to be rethought to minimize
the amount of wasted road space. An external fog-based observer can customize
the lane configurations to safely pack the increasingly heterogeneous traffic
onto the road space while maximizing the packing efficiency.

\section{System Architecture}
\label{sec:sysarc}

In this section, we present an architecture for the edge twin. The edge twin is
hosted on the fogs so that it is accessible with the smallest possible latency
from the vehicles. The edge twin is pervasive (limitless), so vehicles can
access it from anywhere within the area of coverage. The tasks (requests)
launched by the vehicles are processed by the applications running in the edge
twin using data gathered by the edge twin or data pushed by the vehicles.

\subsection{Design Goals and Challenges}

The following design goals and associated challenges were taken into consideration in creating the edge twin architecture.

\noindent{\em\bfseries Low Latency:} In control theory, the impact of the time delay between a plant and its controller is a well-studied problem, which is compensated by the popular Smith predictor~\cite{Natori:cv}. The edge twin is
the controller to the system of vehicles and other objects in the roadways. One
of the primary design goals of edge twin is to minimize latency so that a broad
regime of control applications can be hosted in the edge twin.

\noindent{\em\bfseries Fault Tolerance:} The edge twin is distributed across the fogs such that a fog would cater for the requests from vehicles in its neighborhood. The problem is that when a fog fails, the edge twin functions will become unavailable for a group of vehicles. The challenge is to find fault
tolerance mechanisms that would still yield the same low request response
times. The state-machine replication algorithms~\cite{schneider1990implementing} that are the staple of cloud computing are not suitable for fog computing due to the
relatively high inter-fog communication latencies. Additionally, the fault
tolerance schemes need to gracefully degrade the service (e.g., slow down the
traffic) while recovering from faults and also provide consistent recovery
schemes such that vehicles in a given vicinity are mapped onto the same fog.

\noindent{\em\bfseries Limitless System:} The edge twin must be a limitless
system (i.e., scalable without a performance bottleneck) like the Internet so that it can scale to city levels or beyond.  For the edge twin to be limitless,
the request processing times in the edge twin must be
independent of the number of fog servers that host the edge twin.
The request processing times would have three
components: latency to find a fog for a vehicle at a given location, latency to
ship data and get results from the fog, execution time for the task at the fog.
Therefore, to keep the edge twin limitless, the above three times must be
independent of the number of fog servers.

\noindent {\em\bfseries Multiple Applications:} The edge twin is conceived as
a platform that could host multiple applications. Although in this paper we focus on auto-drive
assist, it should be universal to host other applications that can also benefit from
the real-time world state that is held in the edge twin. For example, battery electric cars need accurate traffic conditions to estimate their energy consumption and prompt the driver to charge the vehicle accordingly. An edge twin based application can be ideal for those purposes. Similarly, edge twin can also host traffic flow optimization algorithms that can use a bird's eye view of the traffic conditions (i.e., physical world view) as input to provide optimal routes.

\noindent {\em\bfseries Shared Responsibility:} The edge twin responds to the
drive assist requests using captured data and extrapolates them using
trajectory predictors. Because there is a chance the physical world conditions
have changed in an unpredictable manner, the responses provided by the edge twin
have to be reinterpreted by the vehicle to create the drive actions. In this
paper, the fog sends a hazard map to the vehicle, which is used in conjunction with
the measurements performed by vehicle's sensors to create the drive actions.  The key
idea is that the ultimate drive actions have inputs from both fog and vehicle.

\subsection{System Design}

\begin{figure}
    \centering
    \includegraphics[width=0.9\columnwidth]{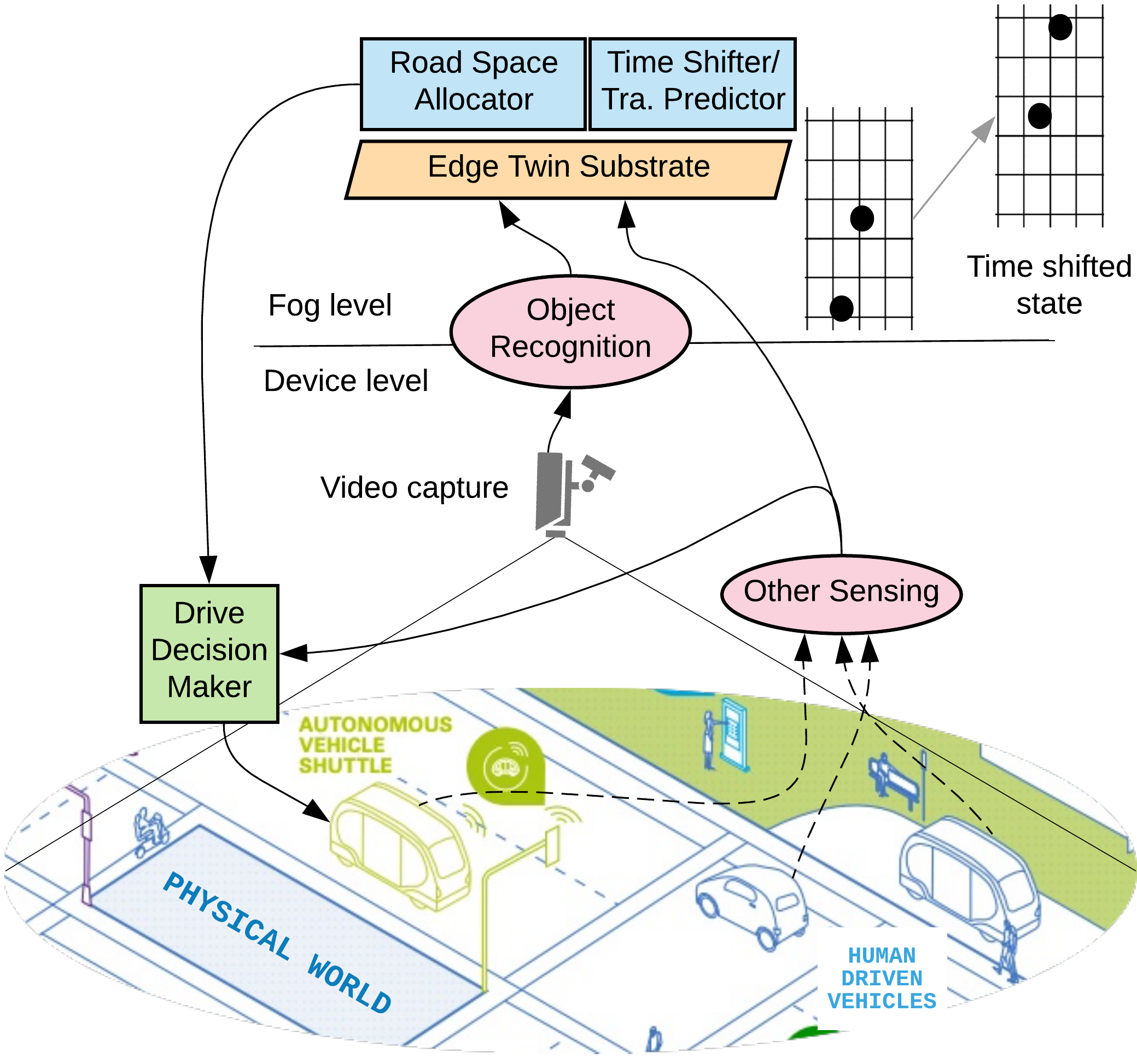}
    \caption{Main task flows of auto-drive assist in the edge twin. Distribution of the edge twin across
    multiple fog servers is not illustrated in this diagram.}
    \label{fig:et_taskflow}
\end{figure}

Figure~\ref{fig:et_taskflow} show the main task flows of the auto-drive assist in an
edge twin for an example scenario. In the example scenario, the autonomous vehicle is sharing the road space with human-operated vehicles, pedestrians, and
cyclists. The happenings in the physical world are captured using an overhead
mounted video camera. An object recognition module that either runs in the
camera or in the fog would process the video to extract the objects and label
them. The edge twin maintains the physical world state by tracking the objects that are relevant to autonomous driving (i.e., a grid of the road space and the
road objects in the grid). The physical world state is continuously updated as
the information feed from the video cameras arrive at the edge twin.
In addition to the camera feed, the individual objects (e.g., cars) can send
GPS coordinate and on-board sensor streams to the edge twin. We need to consolidate all the information
feeds and update the physical world state in the edge twin consistently.

The physical world state in the edge twin is input to the different applications hosted
in the edge twin. The auto-drive assist application uses the physical world state to
provide road space allocations to the vehicles.

The challenge in edge twin task processing is the delay in getting a response from the
edge twin for a mission-critical problem like auto-drive assist compared to an onboard
realization. To mitigate this problem, our architecture time shifts the physical world state and
runs the applications on that state. The edge twin processes the requests from the vehicles based
on this time-shifted world state, which means the vehicle does not need to wait for the full
cycle of processing -- it effectively sees a clairvoyant edge twin.

\subsection{Fault Tolerance in Edge Twin}

\begin{figure}
    \centering
    \includegraphics[width=\columnwidth]{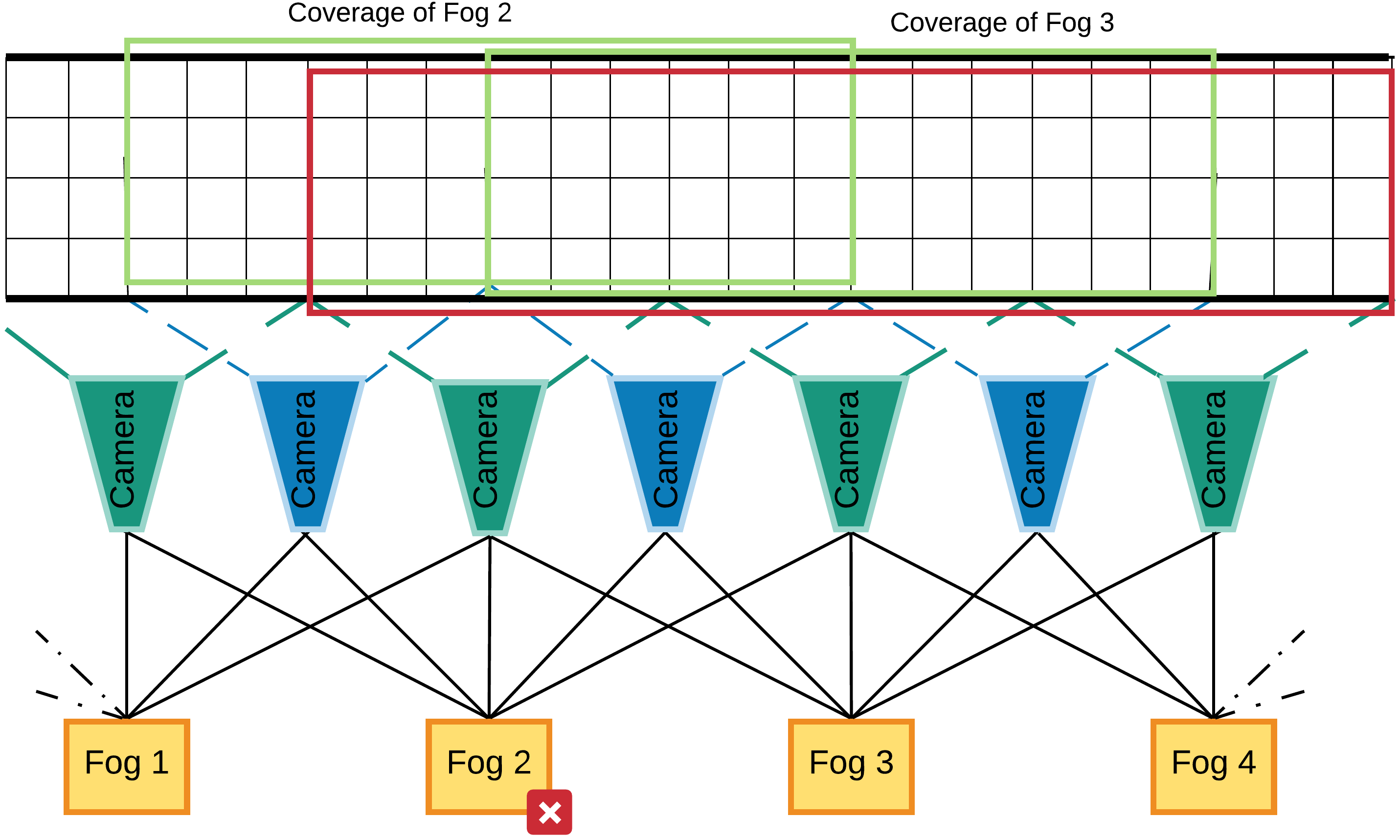}
    \caption{A fault tolerant design for interconnecting fogs and cameras in the edge twin.}
    \label{fig:camera-net}
\end{figure}

The edge twin is a distributed system running across many fogs. Edge twin can fail to respond
the requests from the vehicles, if the fogs that serve the section of edge twin have failed or
the cameras feeding road data have failed.
Figure~\ref{fig:camera-net} shows a fault-tolerant design for interconnecting the fogs and cameras in the
edge twin.

Our design uses two types of cameras: primary (shown in green) and secondary (shown in blue) in
Figure~\ref{fig:camera-net}. The primary cameras provide complete coverage of the road space without
overlapping with each other. The secondary cameras do the same.

The primary and secondary cameras connect to the fogs as shown in the above figure. When there are no faults,
a fog receives feeds from its primary camera and two secondary cameras. This creates an overlap between
the coverages of two adjacent fogs as shown in the green rectangles in Figure~\ref{fig:camera-net}.
The fog can use the coverage overlap to know the traffic conditions upstream and downstream of its
section of the roadway.

Suppose the primary camera in Fog 3 fails, the two secondary cameras would have that covered.
Similarly, if one or both of the secondary cameras fail, the primary camera of Fog 3 and primary
cameras of Fogs 2 and 4 have them covered. The architecture already connects the primary cameras
of Fog 2 and 4 to Fog 3; therefore, secondary camera failures are covered as well.

Now, let us consider Fog 2 failure. To compensate for the failure, Fogs 1 and 3 would enlarge their area of coverage. The
area of coverage of Fog 3 after Fog 2 failure
is shown in a red rectangle in Figure~\ref{fig:camera-net}. Fog 3 is using the feeds from the two adjacent primary
cameras to extend its reach. With the fault tolerance design provided in Figure~\ref{fig:camera-net},
the edge twin would keep working if the failing fogs are not adjacent to each other. In the worst case, up to
half of the fogs could fail and we would still have a working edge twin.

\subsection{Time Shifting in Edge Twin}

\begin{figure}
    \centering
    \includegraphics[width=\columnwidth]{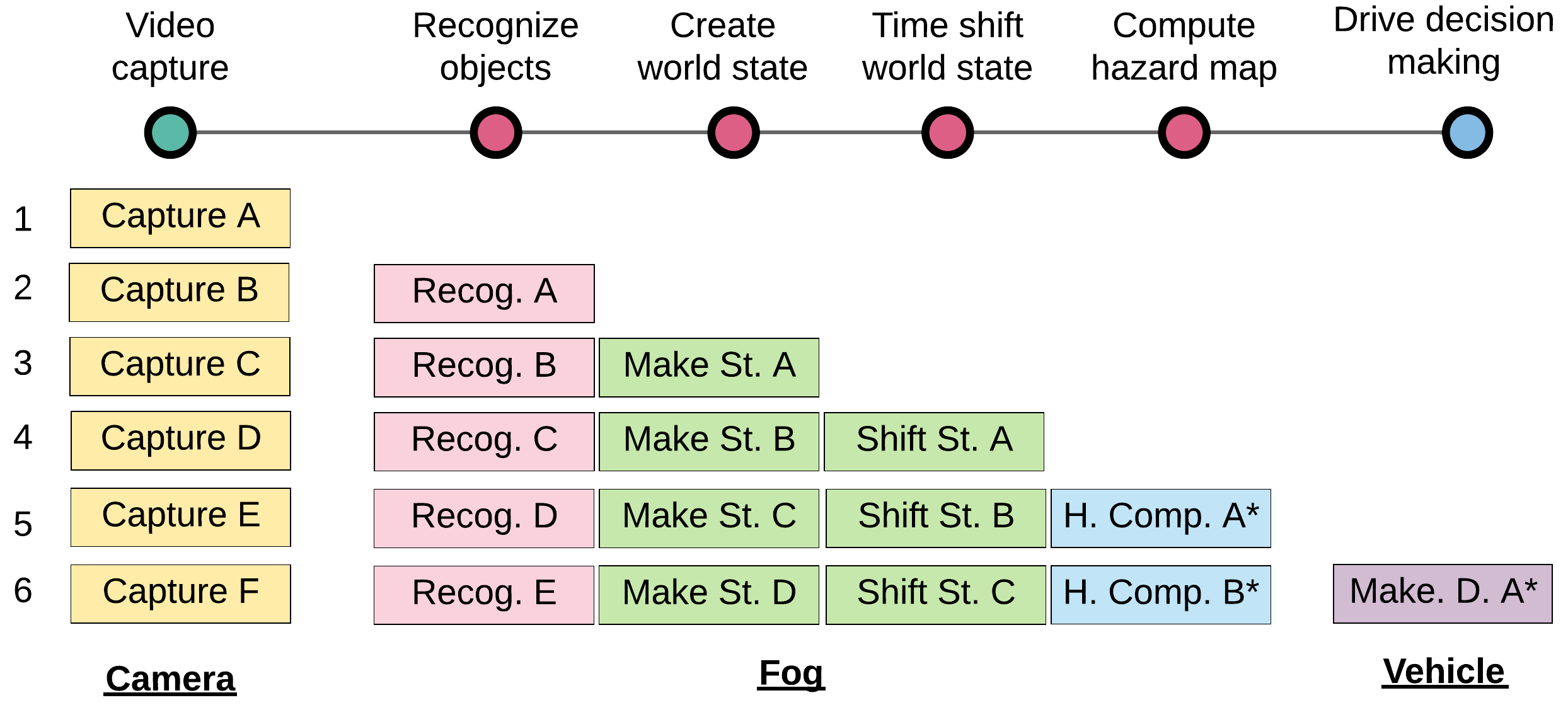}
    \caption{Illustrating the importance of time shifting in the edge twin using the tasks that are in parallel execution at different components of the edge twin at a given time.}
    \label{fig:timeline}
\end{figure}

Figure~\ref{fig:timeline} shows a timeline for the task execution in the resource pipeline from a camera
to a car. Let the system start at time step $1$. The camera executes the {\em capture} task to get scene $A$.
In the next time step, while the camera gets scene $B$, the fog gets the scene $A$ captured in the time step $1$
and runs an object recognition task on it. In time step $4$, the fog runs the time shifting task on scene $A$. The purpose of time shifting can be understood by looking at time step $6$. The system is driving the car in that
time step using data gleaned from scene $A$. However, at that time step, the system is capturing scene $F$. The
time shift is one way for us to compensate for this difference.

The time shifting is achieved by using machine learning algorithms to predict
the trajectory of all the objects in the scene and determine where they would be
at time step $6$. For instance, we could have vehicles, pedestrians, and
cyclists in a scene and we would account for their movement using the machine
learning models. In the ideal case, the time-shifted scene $A^*$ would closely
match the captured scene $F$ at step $6$. In this case, we would have
effectively created a driving directive at time step $6$ from the scene $F$
captured in that step itself, thus hiding the processing latency.

In this paper, we use boosted trees as the machine learning model.
The features for this model are the positions and velocities of the vehicles in the physical world.
The model is trained to predict future positions of the
objects in the edge twin for multiple time steps in the future using features
generated from the objects' motion and trajectory in the past.

The machine learning model brings an additional caveat to the system because
the model tends to perform quite well for immediate future and not so well for distant future.
The accuracy of the model tends to deplete with the number of time steps.
However, long-term predictions are more invaluable in the context of our system than short-term predictions. In addition to improving the long-term prediction accuracy, the choice of using a machine learning model is also influenced by the need to predict a vehicle's location keeping the position of other vehicles in context, which is where conventional kinematics based models struggle. The model can be continuously trained
at the edge twin or even at the cloud (with it global reach) to its accuracy over time.

\subsection{Road Allocation using Box Algorithm}

As shown in Figure~\ref{fig:et_taskflow}, one of the key functions of edge twin is to host applications
that can use the physical world state in the edge twin
to solve queries produced by the vehicles. The road space allocator we briefly describe here is one such
application. The road space allocator considers the road space as a
resource (e.g., like memory in normal operating systems) and finds an allocation for the autonomous
vehicles. The key objective is to find ``free'' space that is {\em highly unlikely} to be used by human-driven vehicles and
allocate them to autonomous vehicles. We also need to ensure that multiple autonomous vehicles do not attempt
to use the same free space and run into each other. The road space allocator uses a box algorithm that discretizes each lane of a stretch of road
into uniform segments called boxes (this idea is used in several papers dealing with auto-driving~\cite{deo2018convolutional}).

A simple box algorithm would continuously track the locations of the vehicles (i.e., using the values
in the edge twin) and map them into the boxes. So, the algorithm would know whether a box is occupied by
a vehicle or not. Using the vehicle trajectory computing module, the box algorithm can know the future
box occupations. When a vehicle reaches out to the edge twin, the box algorithm would determine the
box that holds the vehicle and the projected and/or current occupancy of the boxes that the vehicle is
expecting to use in its future trajectory. Based on a combination of these occupancies, the box algorithm would derive a hazard map that would indicate which box is safe for the vehicle to traverse through in its journey.
By using the vehicle's motion in context with all other vehicles on the road the edge twin can provide
guidance that has an outside view of the traffic conditions.

\section{Implementation}
\label{impl}

We implemented a proof-of-concept simulation of the key portions of our system architecture to gauge its potential performance. This involved choosing a suitable vehicle trajectory dataset, training and evaluating the trajectory prediction model, implementing the box algorithm for road allocation and lastly simulating the road allocation.  

\label{sec:implement}
\subsection{Dataset}
The primary purpose of the proof-of-concept study is to evaluate 
the validity of some of the key ideas relied upon by
our system architecture. We use a trace-driven simulation study and use 
Highway Drone or highD dataset~\cite{highDdataset}. The dataset consists of naturalistic vehicle trajectories on German highways recorded using an overhead drone. It provides a large number of vehicles (110,500) and six different highway locations. The data is presented frame by frame for each vehicle and contains additional valuable information including the locations of surrounding vehicles. For our experiments, we focused on one particular location from the dataset containing trajectories from 1000 cars contained in 22,539 frames. This gave us a combined dataset of 336,186 trajectory points for the vehicles. As standard in machine learning, we divided the dataset into 75/25 train-test split. Some cars from other locations were also used in testing to see the generalization performance of the machine learning models. 

\subsection {Trajectory Prediction}
The trajectory prediction module is an important component of the edge twin to keep results relevant for the vehicles. We use an ensemble of boosted decision trees for the prediction, implemented using the popular Xgboost library \cite{chen2016xgboost} in Python. In the preprocessing phase, each vehicle's past motion is used to create features. The highD dataset already provided data decomposed into Cartesian coordinates and contains features such as the vehicles velocity, acceleration, current lane as well the position of vehicles around it. In addition to this, we also used these metrics from the vehicles historical motion from a sequence of frames in the past to calculate the future position. The target for the model is the displacement of the vehicles (the delta) to a position at a time-step in the future.  
The final model consists of 3000 decision trees with a linear regression objective function to map the output to a set of real value corresponding to the future trajectory of the vehicle. An additional requirement from the model was the ability to predict a flexible number of time-steps in the future. Initially, we trained one model predicting just one time-step in the future and reused its output and as input in the prediction phase to predict multiple time-steps in the future. However, error tended to accumulate with this method for large prediction windows, so we trained different models with different prediction windows in the future. This gave us the luxury of being able to predict a flexible amount of time-steps in the future without compromising on accuracy and therefore safety. 

\begin{figure}
    \centering
    \includegraphics[width=\columnwidth]{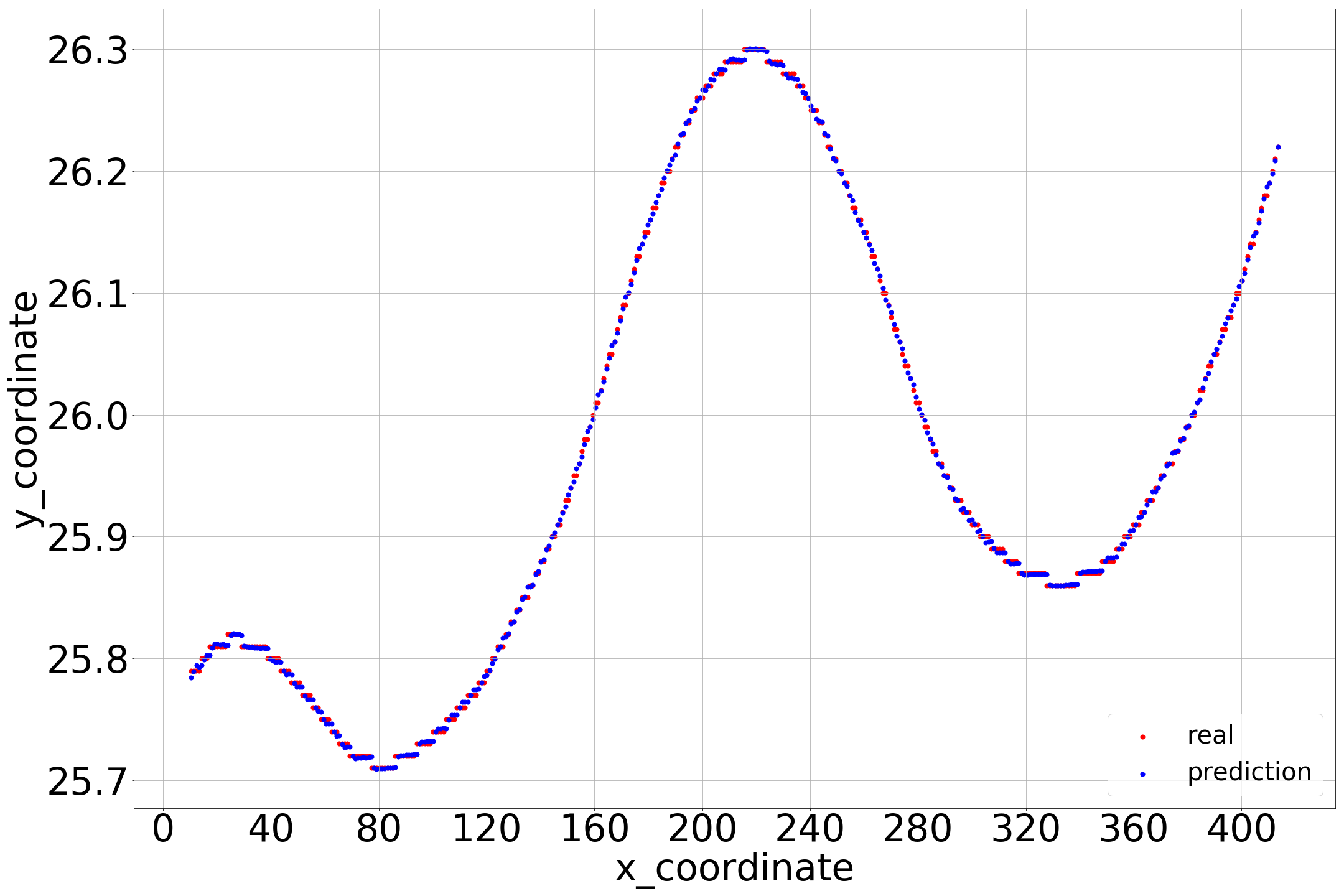}
    \caption{A comparison of the predicted output from the trajectory prediction model with the actual trajectory of a sample vehicle for just the x-coordinate (position along the lane)}
    \label{fig:coordpred}
\end{figure}

\begin{figure}
    \centering
    \includegraphics[width=\columnwidth]{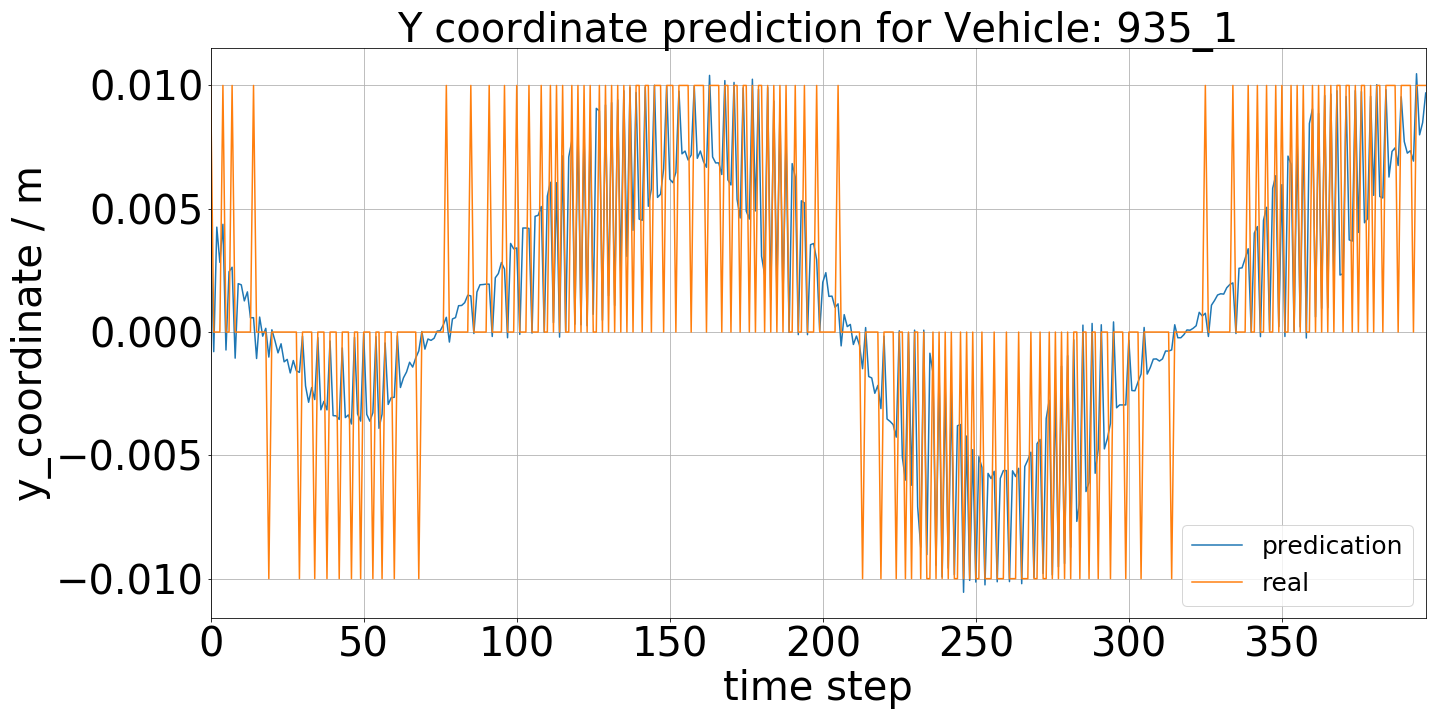}
    \caption{A comparison of the predicted output from the trajectory prediction model with the actual trajectory of a sample vehicle for just the y-coordinate (position in lane).}
    \label{fig:ycord}
\end{figure}

\begin{figure}
    \centering
    \includegraphics[width=\columnwidth]{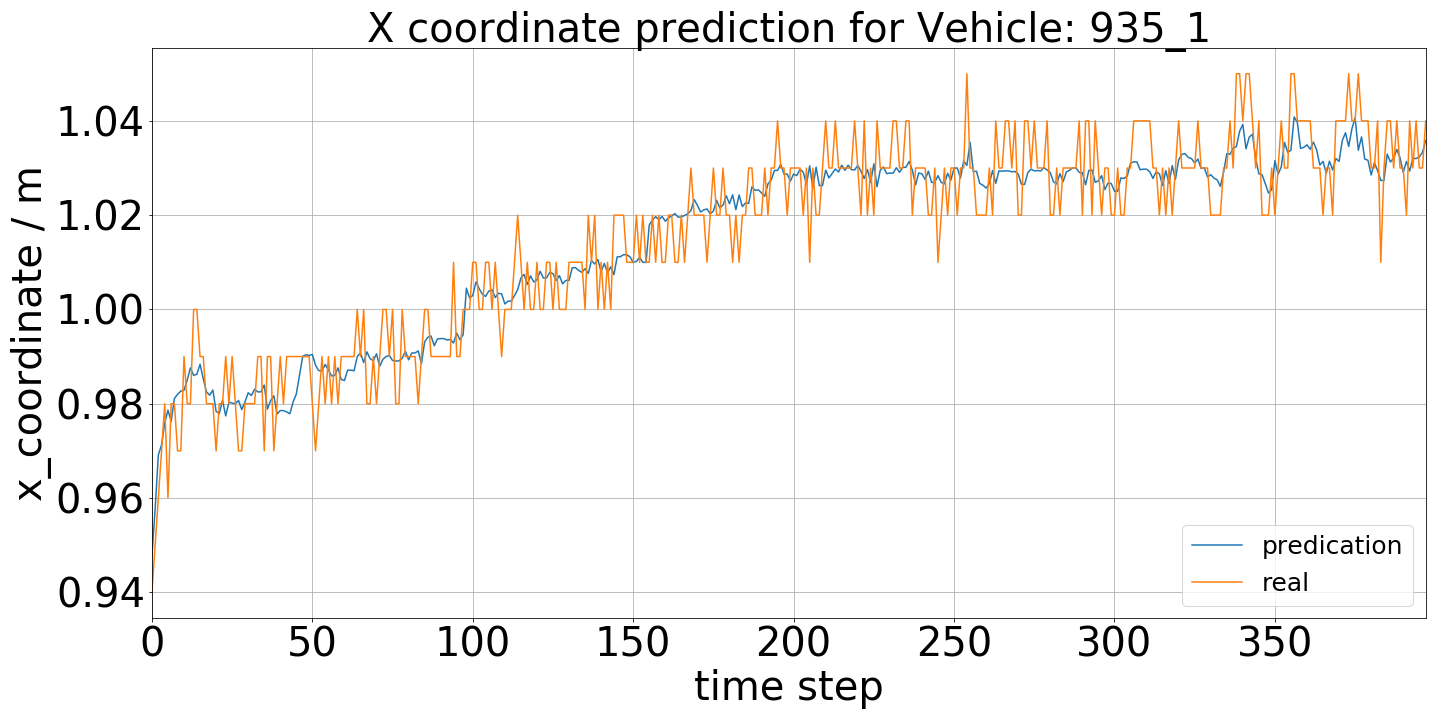}
    \caption{A comparison of the predicted output from the trajectory prediction model with the actual trajectory of the vehicle. This particular vehicle remained well within its as observed by the small variation in its y coordinate. }
    \label{fig:xcord}
\end{figure}

\subsection{Box Algorithm for Road Allocation}
We introduced the box algorithm as a method of allocating each vehicle a safe section of the highway (box) that it could occupy. Each lane is divided into 3 boxes with a length of 5 metres.  For our implementation, we choose the simple criteria of allocating a box that would not result in a collision with any other vehicle. 

The algorithm also utilizes the trajectory prediction module described above to see a view of the world surrounding the vehicle in the future.  A prediction  for 1 second (25 frames) is run for all the vehicles in the vicinity of the vehicle making the request to see which boxes they would be occupying in the future. Moreover, the size and class of the vehicles is also taken into account to see how many boxes they would be occupying. This allows us to generate a hazard map, giving a vehicle a clear idea of which boxes will be safe to occupy in the future and which can be dangerous and potentially cause a collision. 

\subsection{Driving Simulator}

The final section of the framework implementation was designing a decision making algorithm that could direct vehicles to the correct box allocation based on the hazard map provided by the box algorithm. This simulator would output a maneuver such as a lane change decomposed into a series of coordinates the vehicle needs to follow to arrive at the correct box.  

We chose to implement the driving simulator as a convolutional neural network that would take the hazard map for a vehicle as input and the actual position of the vehicle in a future time-step as a target. The key idea here is that we teach the neural network to direct the vehicle in based on the trajectory the vehicle took in the actual trace. This ensures that the vehicle navigates safely to the right box since there are no collisions in the actual traces.  

\section{Experiments and Results}
\label{expr}

\subsection{Performance of Trajectory Prediction Model}

\begin{figure}
    \centering
    \includegraphics[width=\columnwidth]{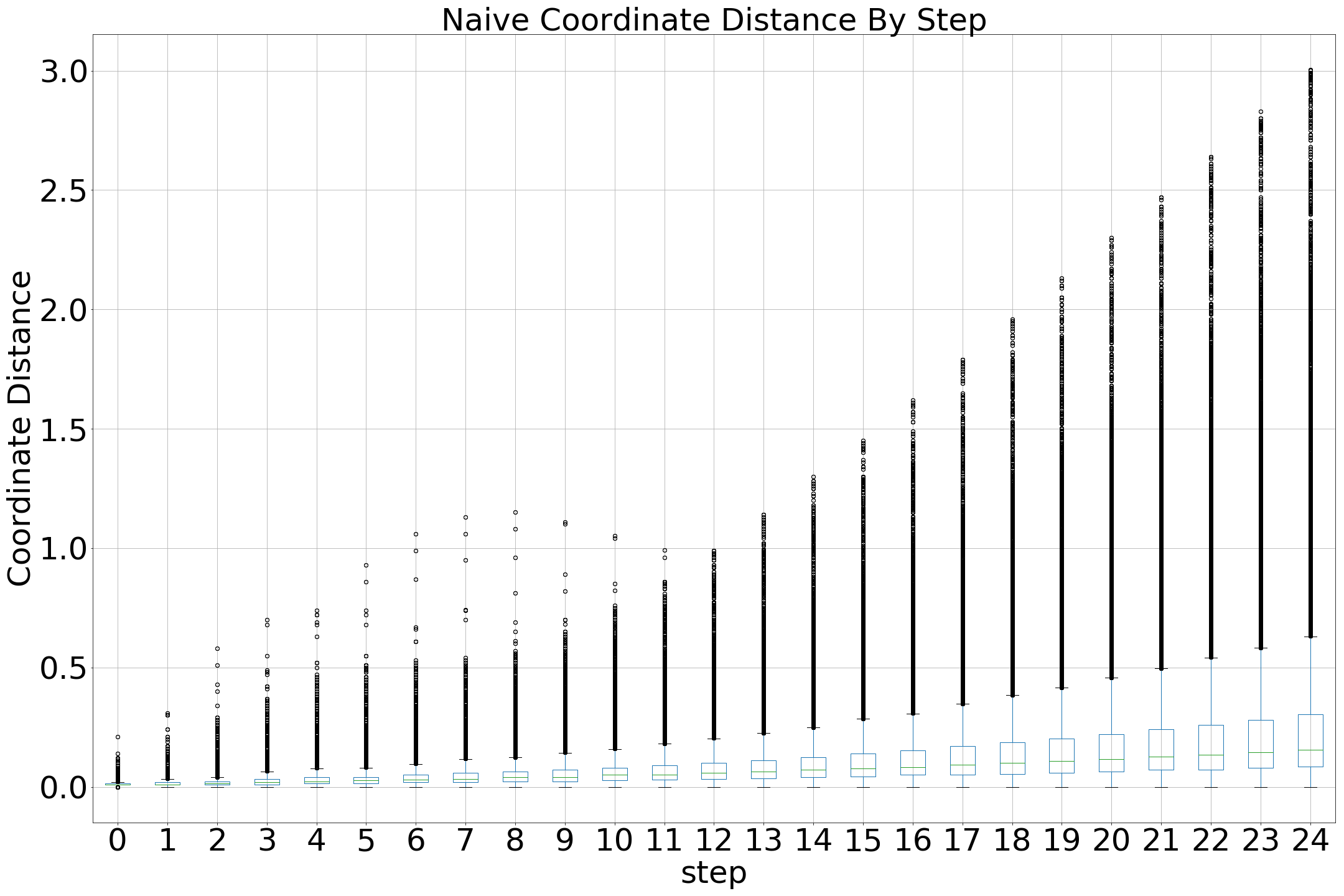}
    \caption{A box plot of the error (called the coordinate distance) between the naive prediction and the actual value. The naive prediction is the value of the coordinates one time-step in the past.}
    \label{fig:naive}
\end{figure}

\begin{figure}
    \includegraphics[width=\columnwidth]{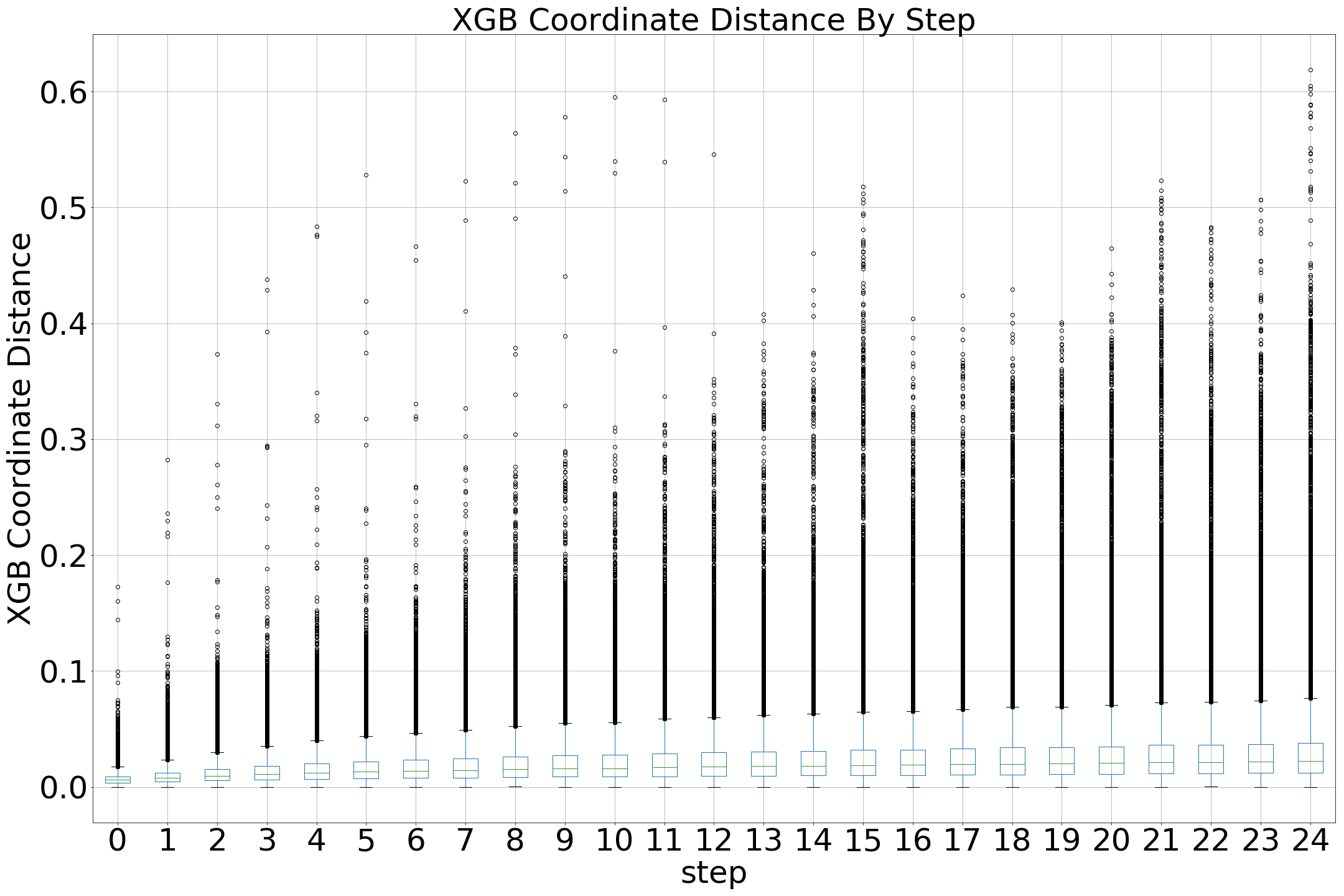}
    \caption{A box plot of the error (called the coordinate distance) between the Xgboost model prediction and the actual value. The Xgboost model performs significantly better than the naive assumption. Please note the different order of magnitude for the vertical axis in this graph. }
    \label{fig:xgboost}
\end{figure}

The trajectory prediction models are one of the most important components of the predictive edge twin. Figures \ref{fig:naive} and \ref{fig:xgboost} show the box plots for the naive prediction and the Xgboost model based prediction. The naive prediction is a popular benchmark for time-series predictions where the last value observed in the time-series is used as the prediction. As obvious in the graphs, the Xgboost model shows a significantly lower coordinate distance for predictions on our test set indicating a much more accurate prediction.

The graphs also depict how the performance varies as we increase the prediction window and predict a large number of time-steps in the future. While the naive prediction's error continues to rise with larger prediction windows, the error for the Xgboost model tapers off. 

Lastly, these graphs show the statistical description of the coordinate distance which is the distance between the predicted value and the actual value for the test set. Although the model was trained using mean squared error as the error metric, this measure gives a more intuitive representation of the prediction for our application.

\subsection{Performance of Speculative Processing}

Successful speculative processing at the fog as described in Section \ref{sec:sysarc} would require a highly accurate prediction to offer any latency gains for autonomous vehicles using this framework. To evaluate this, we evaluated the quality of the prediction for our test set as shown in Table \ref{tab:specfogperf}. We chose a variety of different threshold values for the allowable error between the prediction and the ground truth. Moreover, for these thresholds, we evaluated the performance of three of our 25 trajectory prediction models. Our models showed a high degree of accuracy (above 95\%) in predicting up to 10 frames in the future within an error margin of 5 cm (0.05m). To put this threshold value into context, the width of the lane in this dataset was 3.5m.

\begin{table}
  \caption{Speculative fog processing performance evaluation}
  \label{tab:specfogperf}
  \begin{tabular}{ccl}
    \toprule
    Threshold & Prediction Window Size & Accuracy
    \\
    \midrule
    \multirow{3}{*}{0.5m} & 1 frame & 100\% \\ & 5 frames & 100\% \\ & 10 frames & 100 \% \\ \hline
    \multirow{3}{*}{0.1m} & 1 frame & 99.99\% \\ & 5 frames & 99.81\% \\ & 10 frames & 99.68 \% \\ \hline
    \multirow{3}{*}{0.05m} & 1 frame & 99.98\% \\ & 5 frames & 97.21\% \\ & 10 frames & 95.63\% \\ \hline
    \multirow{3}{*}{0.01m} & 1 frame & 78.17\% \\ & 5 frames & 36.18\% \\ & 10 frames & 29.38 \% \\ \hline
\end{tabular}
\end{table}

\subsection{Transfer Learning}

An additional requirement of the trajectory prediction model in this framework is easy transfer learning to accommodate new fogs merging into the system. To evaluate this, we trained a model on a particular highway section and tested using trajectories of vehicles from a different highway at a different time of the day. Table \ref{tab:transferlearning} shows a comparison of the performance of the model trained on highway 1 and tested on test sets from highway 1 and highway 2.

As expected, the performance on highway 1 is better but the performance on the test set from highway 2 is still quite good and within acceptable margins of error showing that a model the same model can be applied to different sections of the road.

This feature would allow models to be deployed quickly on new fogs. Naturally, the road segments being governed by the fogs should be similar in terms of number of lanes and turns. Moreover, since the system is a continuous learning system, the transferred model would learn the local nuances of its road segment to give a high level of accuracy quickly after deployment.

\begin{table}
  \caption{Transfer learning performance for trajectory prediction model trained trajectory dataset from Highway 1. The difference in the errors is minute indicating that a model trained on one road segment can be successfully applied to another road segment. }
  \label{tab:transferlearning}
  \begin{tabular}{ccl}
    \toprule
    Highway Dataset & Mean Coordinate Distance / m \\
    \midrule
    Highway 1 test set &  0.015033\\
    Highway 2 test set & 0.017480\\
  \bottomrule
\end{tabular}
\end{table}

\subsection{Box Algorithm for Road Allocation}

\begin{figure}
    \centering
    \includegraphics[width=\columnwidth]{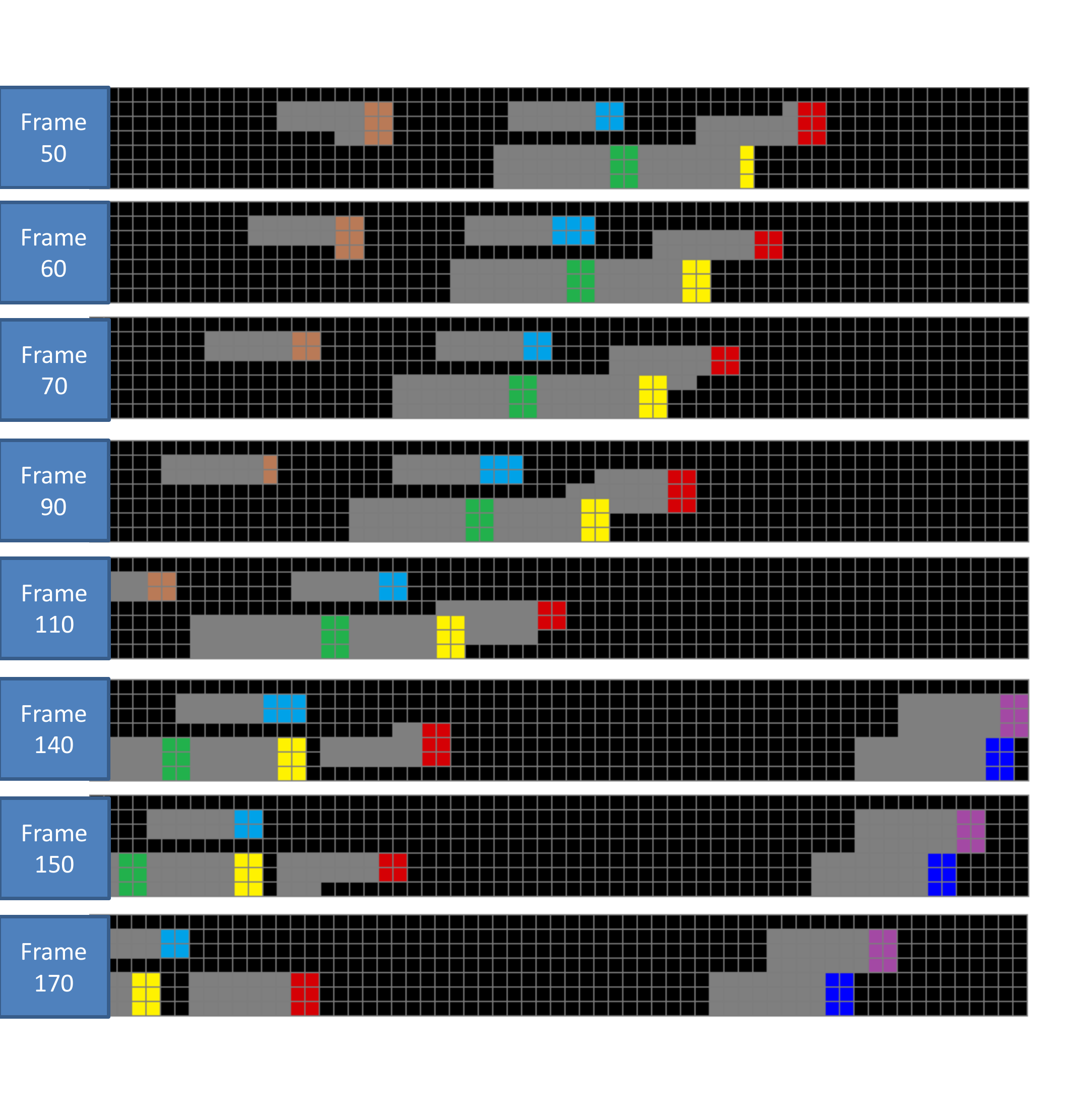}
    \caption{This figure depicts the box algorithm output over multiple frames. Boxes occupied by cars are depicted in colors.  The prediction is computed for all the cars in this scenario and  boxes which would be occupied 6 time-steps in the future in the gray colour. A notable feature is that the algorithm is also able to anticipate lane changes as shown for the car depicted by red boxes in the figure. This particular stretch of highway has two lanes divided into 6 boxes with an additional box representing the hard shoulder. There is a variation in the number of the occupied boxes for vehicles due to differences in size and because vehicles can partially occupy multiple boxes while travelling.  }
    \label{fig:boxalgo}
\end{figure}

Lastly, we gauged the performance of the box algorithm for road allocation according to our implementation. Figure \ref{fig:boxalgo} shows a run-time view of the box algorithm in progress for a sequence of frames captured from the camera. The boxes depicted in gray represent areas of the highway that are expected to be occupied in the near future and are therefore deemed dangerous. This particular sequence of frames is also important as it shows vehicles moving with a high variation in speed. Moreover, one of the vehicles is also shown executing a lane change m maneuver which is successfully predicted by our algorithm.

\section{Related Work} \label{sec:related}
In this section, we explore some of the related research pertaining to the use of edge computing and digital twins in autonomous driving. We also look at some of the existing approaches that employ the use of machine learning for vehicle trajectory modeling and prediction.

\subsection{Edge Computing for Autonomous Driving}

The many advantages offered by the fog computing architecture make it an excellent candidate to support autonomous driving frameworks and systems. This potential is reflected in the related literature in this section that features the use of fog computing for various autonomous driving applications. 

Pi-edge \cite{tang2018pi} is one such framework for autonomous driving that features three important components. The first is a run-time layer designed to run on the different heterogeneous computing systems featured in different autonomous vehicles. The second is a lightweight operating system designed to govern different autonomous driving services and their communications. The last component is an edge-cloud offloader designed to offload compute intensive tasks to the edge cloud dynamically to optimize the vehicles energy usage. In addition to validating the success of their framework in conserving vehicles' onboard batteries, their implementation results verify that offloading compute intensive speech recognition and computer vision based objection recognition tasks not only lowers the edge devices' energy usage while also meeting the latency requirements for real-time operation.   

The authors of \cite{feng2018design} propose a distributed cyber-physical system (DCPS) designed to govern connected and autonomous vehicles. The system features three layers: an information collection layer where the infrastructure sensors are used to collect vital information for the vehicles, a cooperation layer where the information is conveyed to certain vehicles and and individual layer where the vehicle's onboard systems use the information to make decisions and plan their trajectory. They briefly discuss the role of edge computing in making their framework a viable prospect for autonomous driving given the large volumes of data being exchanged between the vehicles and infrastructure and the low latency constraints for transmissions. 

Other research works leverage edge computing for different applications within the autonomous driving domain. \cite{gopalswamy2018infrastructure} present the concept of Special Infrastructure Enabled Traffic Corridors (SIETC) where cars are capable of driving autonomously with the aid of special roadside infrastructure, thereby balancing the liability and responsibility of navigating a vehicle safely between the infrastructure and automotive manufacturer. The infrastructure would be responsible for providing and distilling Situational Awareness (SA) to the cars through edge computing to enable this. Another framework \cite{peng2018sdn} proposes the use of an Software Defined Network (SDN) and MEC that disseminates high definition maps of the road environment to vehicles for safer navigation. 

An important distinction between these frameworks and our proposed edge twin framework is that our framework is designed to have an up-to-date global view of the driving world surrounding the vehicle. Moreover, with the predictive capabilities, our framework is also designed to further improve the apparent latency between the infrastructure and the vehicle, to the point where the fog augmented infrastructure is capable of driving an autonomous vehicle in real-time. This is in stark contrast to the approaches we discussed here which propose selectively offloading certain computationally intensive tasks to the edge servers.

\subsection{Digital Twin for Autonomous Driving}

As introduced in Section \ref{sec:background}, research related to digital twins are primarily concerned with the product design and manufacturing industries. However, some works such as \cite{kumar2018novel} propose the use of digital twins for vehicles and roads which can then be leveraged by machine learning based approaches to predict and minimize congestion on roads. The author's of \cite{chen2018digital} also provide an interesting use-case for digital twins as digital behavioral twins designed to learn driver behaviour through data accumulated from connected smart vehicles. The behavioral twins are utilized to compute the risk associated with surrounding cars and recommend (or autonomously take) preventive actions. 

To the best of our knowledge, our proposed edge twin is the first proposed framework that features a distributed digital twin. Moreover, our edge twin design is unique given the large amount of data it will track in a virtual space as well as the ability to predict future views using the in-built prediction module.

\subsection{Trajectory Prediction using Machine Learning}
Recent developments in machine learning have played an important role in the development of autonomous driving \cite{huval2015empirical}. Trajectory prediction for vehicles in naturalistic driving environments is one such application of machine learning in the autonomous driving domain which has attracted a lot of research. A wide range of models exist for vehicle trajectory prediction and have been covered succinctly in \cite{survey}. The Trajectory prediction module is also a key component of our predictive edge twin. In this subsection, we look at some of the approaches taken in literature to predict vehicle trajectories.

The authors in \cite{deo2018multi} utilize an LSTM based model to predict the future trajectory of a vehicle keeping the motion of surrounding vehicles in context. The encoder-decoder model takes the previous track histories of each vehicle as inputs and outputs the distribution of future vehicle locations as parameters of a Gaussian distribution, effectively providing probabilities for road occupancy. A secondary model for also utilizes the same inputs and is utilized for maneuver  prediction. This model outputs the maneuver specific probability distributions with 6 possible maneuvers considered in the study. This paper utilizes the publicly available NGSIM US-101 and I- 80 datasets. Similar to our dataset, these datasets are also based on video footage from overhead cameras. 

The same authors build on their existing approach in \cite{deo2018convolutional} by adding convolutional social pooling into their previous LSTM based model. The social context of surrounding vehicles is modelled using a convolutional neural network and the output concatenated with the LSTM model of the ego vehicle to accomplish the same trajectory and maneuver prediction and performs better in terms of prediction error than other approaches. 

In contrast to overhead views, trajectory prediction can also be done using cameras on-board the vehicles. The work done by the authors in \cite{deo2018would} utilizes these cameras to track surrounding vehicles and projects them onto the ground plane. Then, trajectory and maneuver prediction models are run for each of the surrounding vehicles to come up with the most likely trajectories and maneuvers the vehicles would execute. The authors utilize a Hidden Markov Model (HMM) for maneuver prediction using 10 different classes of maneuvers and an ensemble of Bayesian filters for trajectory prediction from the interacting multiple model (IMM) framework. The system is designed to run on-board a car, allowing it to analyze and predict the motion of surrounding vehicles.  
\section{Conclusion and Future Work}

This paper presents a new fog computing based framework called edge twin for auto-drive assist. We established five design goals for the edge twin. One of the key design goals of edge twin is
compensating for the delay of outsourcing the auto-drive assist function to the fog. Without the delay compensation, the responses provided by the edge twin for auto-drive assist
requests would be stale -- they would relate to an older world view. As a result, the vehicles would not be able to use the responses to create their driving decisions.

We develop a trajectory predictor for the vehicles and use it to time shift the world state to a future point. We use the time-shifted world state to create driving decisions that remain
relevant to the vehicles when they receive the decisions. We tested the trajectory predictor using real traffic traces and show that our trajectory predictor is a good mechanism for time-shifting the world state. We discretized the road space in our world state representation and created a road hazard computation procedure. The vehicles receive the hazard maps that are relevant for the time point they are at and can use the maps to form the driving decisions.

Future work will address the remaining design goals and challenges and develop a full-featured edge twin to implement auto-drive assist. The edge twin framework we proposed in this paper will be built using JAMScript -- a programming language for edge-oriented mobile IoT that we have already developed~\cite{wenger2016programming,DBLP:journals/corr/abs-1906-09962}. Several design goals of the edge twin align nicely with those of JAMScript making it a natural host for the edge twin.

\bibliography{refs}
\bibliographystyle{ieeetr}

\end{document}